\documentclass[letterpaper,conference,final]{IEEEtran}

\usepackage[binary-units=true]{siunitx}
\DeclareSIUnit \dBm {dBm}
\DeclareSIUnit \dB {dB} 
\DeclareSIUnit \dBi {dBi} 
\DeclareSIUnit \Kbps {Kbps}
\DeclareSIUnit \Mbps {Mbps}
\DeclareSIUnit \Gbps {Gbps}
\DeclareSIUnit \kBps {kBps}
\DeclareSIUnit \MBps {MBps}
\DeclareSIUnit \GBps {GBps}

\usepackage{cite}      
\usepackage[caption=false, font=footnotesize]{subfig}
\usepackage[dvips]{graphicx}
\usepackage{url}       
\usepackage{amsfonts}
\usepackage{amssymb}
\usepackage{mathtools}
\usepackage{times}
\usepackage{algpseudocode}
\usepackage{fourier}
\usepackage{algorithm}
\usepackage{enumerate}
\usepackage{multirow}
\usepackage{tikz}
\usepackage{array}
\usepackage{gensymb}
\usepackage{eucal}
\usepackage{mathtools}
\usepackage{nicefrac}
\usepackage{hyperref}
\usepackage{tabularx}

\newcolumntype{P}[1]{>{\centering\arraybackslash}p{#1}}

\bstctlcite{IEEEexample:BSTcontrol}

\algnewcommand\algorithmicinput{\textbf{Level 1:}}
\algnewcommand\Level{\item[\algorithmicinput]}

\algnewcommand\algorithmicinputt{\textbf{Level 2:}}
\algnewcommand\Levell{\item[\algorithmicinputt]}

\algnewcommand\algorithmicinputtt{\textbf{Level 3:}}
\algnewcommand\Levelll{\item[\algorithmicinputtt]}

\algnewcommand\algorithmicinputttt{\textbf{Output:}}
\algnewcommand\Output{\item[\algorithmicinputttt]}

\usepackage[english]{babel}
\usepackage{blindtext}
\begin{document}
\title{Poster: Parallel Implementation of the OMNeT++ INET Framework for V2X Communications}
\author{\IEEEauthorblockN{Ioannis Mavromatis, Andrea Tassi, Robert J. Piechocki, and Andrew Nix}
  \IEEEauthorblockA{Department of Electrical and Electronic Engineering, University of Bristol, UK \\ Emails: \{Ioan.Mavromatis, A.Tassi, R.J.Piechocki, Andy.Nix\}@bristol.ac.uk}
}

\maketitle

\begin{abstract}
The field of parallel network simulation frameworks is evolving at a great pace. That is also because of the growth of Intelligent Transportation Systems (ITS) and the necessity for cost-effective large-scale trials. In this contribution, we will focus on the INET Framework and how we re-factor its single-thread code to make it run in a multi-thread fashion. Our parallel version of the INET Framework can significantly reduce the computation time in city-scale scenarios, and it is completely transparent to the user. When tested in different configurations, our version of INET ensures a reduction in the computation time of up to $43$\%.
\end{abstract}

\section{Introduction and Motivation}
\vspace{-2mm}
Recently, more and more resources are being allocated to developing of sophisticated communication frameworks for the next-generation Intelligent Transportation Systems (ITSs). As such, the issue of running cost-effective city-scale experimentation is often addressed by means of simulation frameworks~\cite{8275604,10.4108/eai.20-3-2018.154368}. Unfortunately, existing simulation frameworks (i.e., ns-2, INET, etc.) execute in a single-thread fashion. Thus, in city-scale scenarios, one minute of simulation can easily result in days of computation.

Researchers in the area of Parallel Discrete Event Simulations (PDES) have tried to leverage from new high-performance computing platforms for a very long time. Unfortunately, this resulted in parallel features that require a quite laborious reconfiguration of the existing scenarios, without always assuring improvement in the simulation time~\cite{omnetpp}. For these reasons, existing simulation frameworks are mainly operated in a single-thread fashion. The necessity for parallel models, motivated us to investigate the existing frameworks and exploit ways of parallelizing them. In this work, we investigate the INET Framework~\cite{inetFramework} that is an open-source library for OMNeT++ simulation environment~\cite{omnetpp}. INET is one of the most well-known tools for vehicular simulations being a full-stack network simulation framework. In this poster, we will identify the sequential functions of INET related to the exchange of wireless packets and analyze the way they can be parallelized. Then, we will present our multi-threaded version of the aforementioned functions. Our multi-thread implementation of the INET Framework can be downloaded from {\tt\small https://github.com/v2x-dev/multithread-inet}.

\section{System Analysis and Proposed Solution}
\vspace{-3mm}
We identify at first the way INET interprets obstacles and propagated signals. In a vehicular scenario, buildings are the main obstacles that can be found in a city. INET refers to them as obstacles and parses them within the \emph{PhysicalEnvironment} namespace. All the obstacles are listed in an XML file with an attribute \emph{type}, which defines their shape. Each obstacle is represented by a set of coordinates, that are regarded as the edges of the building in the 3D space. Each building is also associated with a specific \emph{material} that is being used to calculate the attenuation loss caused by the obstacle. The obstacle loss in INET is calculated by two different models, the \emph{IdealObstacleLoss} and the \emph{DielectricObstacleLoss}. The first determines either the signal as completely blocked, or not attenuated when intersected  with a physical object. 
The latter computes the power loss based on the material properties, the shape, the position and the orientation of an obstacle.

INET treats the positions of each vehicle as a point on the simulation canvas and updates them by using the SUMO traffic generator~\cite{sumo}. The signal propagation is modeled as a line segment, between point A and point B. 
The signal attenuation is a function of: 1) the path loss model, 2) the obstacle loss model from the number of intersections calculated as mentioned before. Both models are configured by the user.
What is really of interest, is the way INET calculates the attenuation due to the wall intersections. When a packet is transmitted, INET finds in a sequential and iterative manner all the intersections with the obstacles. For the given intersections, it calculates the obstacle loss using the function \emph{visit} of the obstacles classes mentioned above.
Considering the above and that IEEE 802.11p operates in broadcast mode, it is evident that the above process is computationally expensive. In fact, the computation of the attenuation loss has a computational complexity of $O(mn^2)$, where $n$ is the vehicles and $m$ the number of obstacle intersections. This significantly increases the simulation time in large-scale scenarios.

In order to overcome the aforementioned problem, we developed a multi-thread version of the \emph{PhysicalEnvironment} class by modifying the function \emph{visitObjects}. This function is responsible for parsing all the obstacles and finding the power attenuation. Also, we modified \emph{visit} in \emph{DielectricObstacleLoss}, to ensure flawless operation. In our version of INET, the number of threads can be dynamically changed by the user when initializing a scenario.

In city-scale scenario (namely, maps greater than $\geq \SI{2}{\kilo\meter}$), vehicles are not expected to communicate from one side of the city to the other. Despite this, INET always computes the intersection map between each pair of vehicles and their signal attenuation, regardless of the distance between them. In order to speed up the execution time even further, we integrate the notion of the transmission radius in the system. As such, we introduced the \emph{distanceBoundary} user parameter, within the \emph{ScalarAnalogueModel} class, under the \emph{RadioMedium} namespace. For all the exchanged packets, we find the distance between the two communicating vehicles, and if it is greater than the given boundary, we regard the packet as non-deliverable. By that, we can avoid unnecessary calculations in the \emph{PhysicalEnvironment} model. Of course, the above improvement can be implemented into other analog models as well (e.g., \emph{DimensionalAnalogModel}) but we chose the scalar one as a proof of concept.

\begin{table}[t]
\renewcommand{\arraystretch}{1.02}
\centering
\vspace{-5mm}
    \caption{List of Simulation Parameters.}
    \begin{tabular}{r|l||r|l}
    \textbf{Parameter} & \textbf{Value} & \textbf{Parameter} & \textbf{Value} \\ \hline \hline
    Simulation time & \SI{100}{\second} & Carrier Frequency & \SI{5.9}{\giga\hertz} \\
    TX Power & \SI{25}{\dBm} & Channel Bandwidth & \SI{10}{\mega\hertz} \\
    TX/RX Antenna Gain & \SI{9}{\dBi} & Message Length & \SI{140}{\byte} \\
    RX Sensitivity & \SI{-93}{\dBm} & Pathloss Exponent & 2.4 \\
    Cable/System Loss & \SI{3}{\dB} & Distance Boundary & \SI{1000}{\meter} \\
    Transmission Interval & \SI{0.1}{\second} & & \\  \hline
	\end{tabular}
\label{tab:parameters}
\end{table}

\vspace{-2mm}
\section{Performance Evaluation and Discussion}
We evaluate the performance of our implementation with two large-scale scenarios in a grid-like fashion. The simulation parameters are as shown in Table~\ref{tab:parameters}. At first, we evaluate the execution time as a function of the number of vehicles for a map of size $\SI{2}{\kilo\meter}^2$. Then, we present the execution time as a function of the map size. We consider six map sizes $\left\lbrace \SI{800}{}, \SI{1100}{}, \SI{1400}{}, \SI{1700}{}, \SI{2000}{}, \SI{2300}{} \right\rbrace\SI{}{\meter}^2$ and $100$ vehicles for each scenario. All scenarios have roads equally spread horizontally and vertically every \SI{100}{\meter}, without traffic lights at the intersections. Each road is $2$-lanes wide. Within each road square, we generated buildings with sides of $\SI{950}{\meter}$ that act as obstacles in our scenarios. Finally, for all scenarios, we generated the vehicle traffic by using SUMO~\cite{sumo}.

\begin{figure}[t]
\vspace{-4mm}
\centering
\includegraphics[width=0.95\columnwidth]{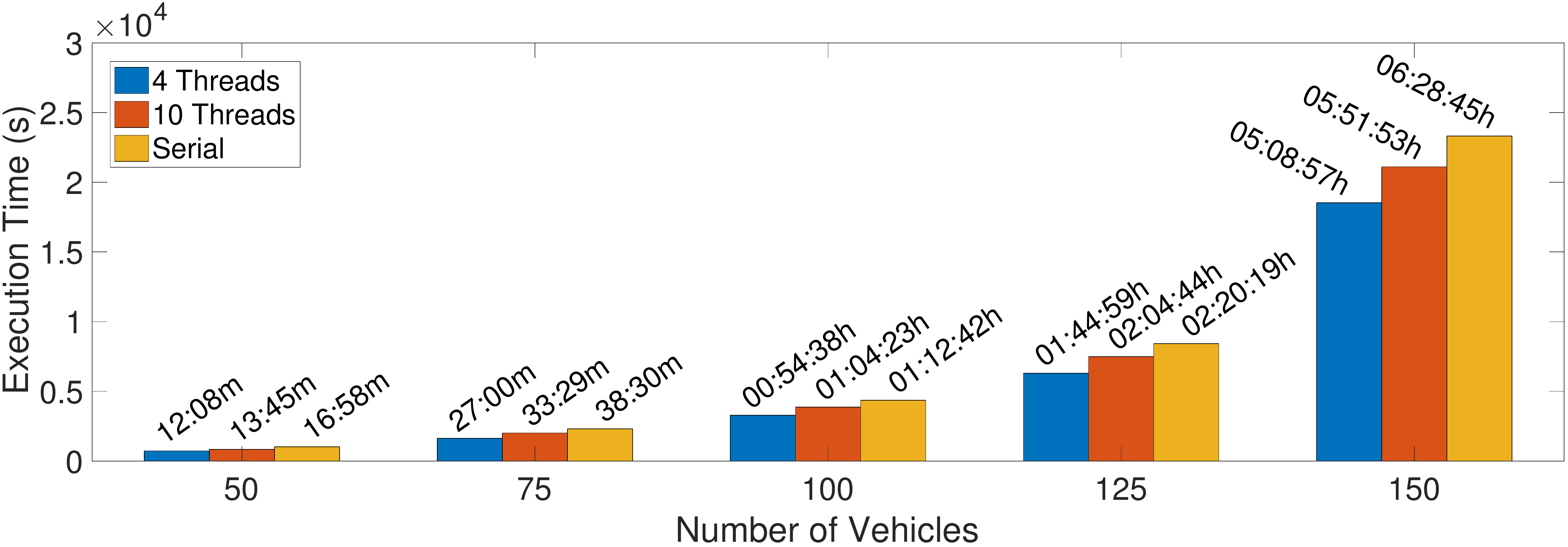}
    \vspace{-3mm}
    \caption{The execution time, measured as a function of the number of vehicles for the parallel and the sequential implementation.}\vspace{-2mm}
    \label{fig:barPlot}
\end{figure}

Fig.~\ref{fig:barPlot} shows the execution time measured as a function of the number of vehicles. Any multi-threaded execution generates some computational overhead in order to create the threads and handle the content-switch. To that extent, we compared two different number of threads ($4$ and $10$). Overall, by parallelizing the functions mentioned above, our version of INET ensures reduced computation times. When four threads are used, we observe an improvement of up to $30$\%. When ten threads are utilized, the improvement ranges between $10\%$ and $12$\%. The increased overhead of the number of threads in the second case is the reason for this difference in the performance. We observed that the optimum number of threads is scenario dependent and is related to the average number of intersections between the communicating vehicles (one thread per intersection).

\begin{figure}[t]     
\centering
\includegraphics[width=0.95\columnwidth]{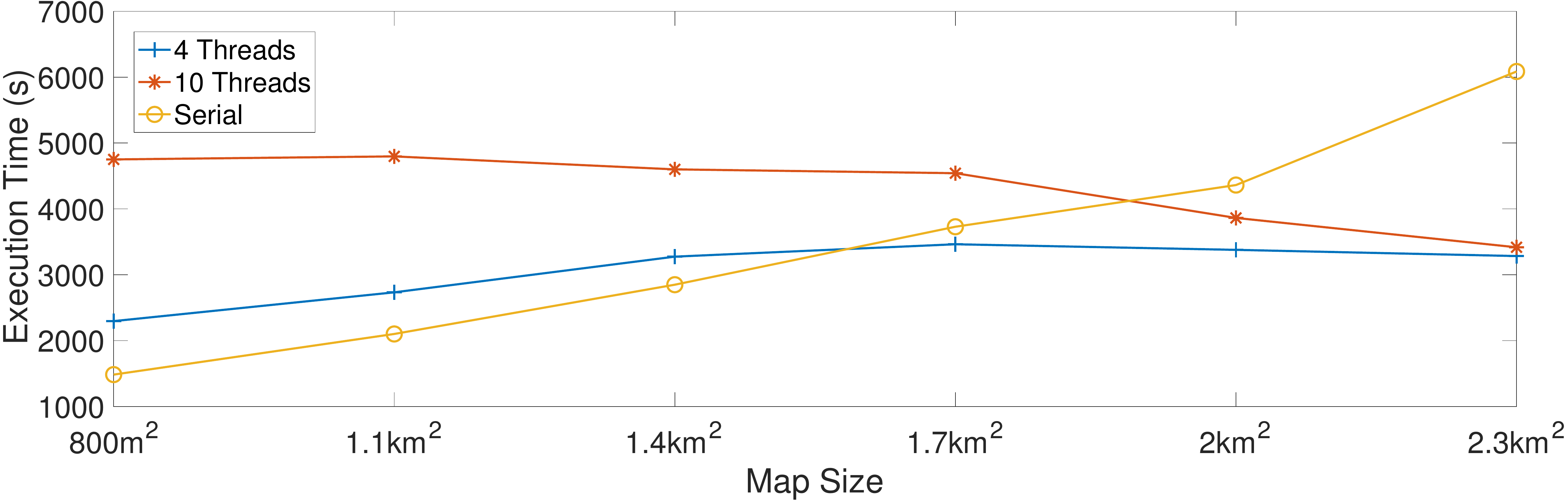}
    \vspace{-3mm}
    \caption{The execution time, measured as a function of the map size for the parallel and the sequential implementation.}
    \label{fig:mapSize}
\end{figure}

Fig~\ref{fig:mapSize} shows the execution time required, as a function of the map size, for $100$ vehicles. The synthetic maps we generated have a relatively small number of obstacles compared to a real city. We observe that for small maps (from $\SI{800}{\meter}^2$ to $\SI{1.4}{\kilo\meter}^2$) the sequential INET achieves slightly better performance compared to the parallel one. This is caused by the multi-thread overhead. However, for larger maps, we observe that our multi-thread version of INET outperforms the sequential one by reducing the computational time of $43\%$, for a $\SI{2.3}{\kilo\meter}^2$ map. Again, when we increase the number of threads, the increased overhead leads to increased simulation time compared to the 4-thread scenario, but still manages to outperform the sequential execution. Finally what we observe is that for maps $\geq \SI{1.7}{\kilo\meter}^2$, as the size of the map increases, the execution time decreases. This is because of the distance boundary that we introduced. For a fixed number of vehicles that are equally spread on the surface of the map, the distance between them will be greater when the size of the map is increased.

\vspace{-1mm}
\section{Conclusions}\label{sec:conclusions}
\vspace{-2mm}
In this work, we investigated the bottleneck of the INET Framework and proposed an optimized multi-thread re-factoring of its code. With our solution, the computation time can be decreased by up to $43\%$ compared to the single-thread version. Our multi-threaded implementation ensures the seamless integration with the existing simulation scenarios of a user and is easily configurable to speedup the simulation time when required. 



\vspace{-2mm}

\bibliographystyle{IEEEtran}
\bibliography{bib.bib,IEEEabrv}

\end{document}